# Urban Analytics: History, Trajectory, and Critique[1]

Geoff Boeing[2], Michael Batty[3], Shan Jiang[4], Lisa Schweitzer[5]

**Abstract.** Urban analytics combines spatial analysis, statistics, computer science, and urban planning to understand and shape city futures. While it promises better policymaking insights, concerns exist around its epistemological scope and impacts on privacy, ethics, and social control. This chapter reflects on the history and trajectory of urban analytics as a scholarly and professional discipline. In particular, it considers the direction in which this field is going and whether it improves our collective and individual welfare. It first introduces early theories, models, and deductive methods from which the field originated before shifting toward induction. It then explores urban network analytics that enrich traditional representations of spatial interaction and structure. Next it discusses urban applications of spatiotemporal big data and machine learning. Finally, it argues that privacy and ethical concerns are too often ignored as ubiquitous monitoring and analytics can empower social repression. It concludes with a call for a more critical urban analytics that recognizes its epistemological limits, emphasizes human dignity, and learns from and supports marginalized communities.

## Introduction

Urban planners and policymakers have long relied on quantitative studies of city patterns and processes to predict and plan for future development. But in recent years, a new field of urban analytics has emerged to answer urban questions in new ways by synthesizing computer science, statistical physics, and machine learning (Batty, 2019). Canonical definitions are tricky in emerging fields, but, in general, *urban analytics* is a broad catch-all term for quantitative workflows for gathering, processing, and analyzing data in a spatiotemporal context that applies statistics and computer science to urban questions. Today the terms *urban data science* and *urban analytics* are frequently used interchangeably (Kang et al., 2019).

The field of urban analytics crosscuts most of the other chapters in this book through the urban dimension. Traditionally, it has focused on deductive methodologies, operating within prevailing theoretical frameworks to make specific predictions. But in recent years, urban analytics has drifted toward more inductive methodologies—foregrounding the data themselves and generalizing new theory from those observations—especially through the empirical lens of spatiotemporal big data and machine learning. Today its key challenge is to move beyond the low-hanging fruit of *analysis* to contribute more broadly and critically to urban theory, real-world policy, and the wicked problems—such as inequality and injustice—that plague city planning (Batty, 2019).

The recent explosive growth of urban data production has fueled this field's rapid expansion. Researchers and practitioners today have access to a cornucopia of spatiotemporal big data—much of it user-generated or passively sensed—that can be computationally modeled and analyzed to generate insights into cities and citizens and to build predictive tools and dashboards for monitoring and guiding urban processes (Kontokosta, 2018). In turn, new university research institutes and degree programs have arisen around the world to focus on urban analytics. But looking beyond this rapid rise, what is the field focused on and does it really make society any better off?

This chapter explores the unfolding trajectory of urban analytics. It begins with a synoptic history of its origins and evolution as a deductive discipline that studied urban systems primarily through equilibrium/aggregate spatial models before shifting toward disaggregation and microsimulation. The remainder of the chapter then reflects on the field's emerging and increasingly inductive trends. First, we discuss spatial network science in the context of cities and their transportation systems. Second, we trace the inductive turn toward spatiotemporal big data and machine learning and review their place in urban practice today. Third, we critically consider problems with past and current approaches to surveilling and quantifying "the urban" and propose ways in which urban analytics can move beyond top-down monitoring to promote emancipatory social transformation and justice.

## Origins of Urban Analytics: Early Theories and Models

### Competing Epistemologies

Formal methods to analyze urban phenomena grew out of classical physics and energy exploiting technologies during the Industrial Revolution. A long line of urban theory developed from these ideas, from the French Physiocrats of the 18[th] century to the German location theorists of the 19th (Isard, 1956). Much of this methodology and theory has since been encapsulated in social physics and its application to transportation modeling from the mid-20th century onwards. The field of statistics emerged in the early 20th century, originally in biology and genetics before expanding across a wide swath of the social sciences. This expansion included a major endeavor to adapt statistics from probability theory to spatial, geometric, and locational problems emphasizing movement, diffusion, population density, and spatial correlation. This fusion of ideas developed into the domain that is now called urban analytics.

Focusing on location theory, social physics, and multivariate statistics, the domain's momentum grew as the digital computer powered novel applications to problems of spatial and economic planning. This emphasized predicting cities through deductive modeling and simulation. It stood in sharp contrast to scientific endeavors based on induction using statistical methods to search for meaningful spatial patterns in large data sets. To some extent, this methodological chasm marked the different goals between professional and academic communities developing urban analytics, with the former focusing on real-world planning and policy while the curiosity-driven latter attempted to *explain* rather than *redesign* spatial patterns.

Until the late 20th century, urban analytics focused on simulation models for prediction. But over the past 30 years it has shifted toward explaining spatial structure and dynamics through statistical models, due to both the availability of new digital data and the growing dissatisfaction with the dominant paradigm of scientific deduction. Induction has instead grown more prominent as the quest for building statistical space-time models reorients itself toward multivariate machine learning methods. This shift has coincided with the rise of big data, largely from embedded and mobile sensors which now generate massive volumes of spatial data in real time. It also coincides with a retreat from attempting to build predictive models of the long-term future of cities, partly due to continued wrestling with the notion that urban science will never be able to accurately predict the distant future due to massive complexity and interdependence. All it can do is inform wider debates about what the future could be like.

**Evolution of Urban Models**

The models that first dominated urban analytics were based on deterministic structures representing how different urban spatial components related to one another through traffic, migration, economic linkages, aging, etc. These transportation-land use models embodied flow relationships and articulated them using concepts from social physics, gravitation, input-output analysis, economic utility and decision theory (de la Barra, 1989). They contrasted with simpler econometric models that used only rudimentary theories of spatial systems, but were specified in probability terms and focused on causal mechanisms. A long line of space-time models, some of which arose in urban policymaking, established this tradition seeking to generate consistent estimators whose statistical properties were well-defined.

These early models usually treated spatial systems as though they were in equilibrium. But as researchers learned more about these systems and developed better methods of fitting models to data, they developed more complex models with explicit dynamic processes. Many of these early approaches could not be easily generalized, disaggregated to finer scales (e.g., individual objects or attributes), or made dynamic. Newer models however began to focus on development processes—for example, cellular automata models and agent-based models for demographic forecasting, traffic and pedestrian movement, and migration (Crooks et al., 2019). Variants, such as those based on microsimulation that focused on simulating integrated microdata, also appeared.

Many of the urban models still in use today have their origins in the earliest applications. particularly in large-scale residential and housing market applications such as the UrbanSim modeling framework (Waddell, 2002) as well as in travel demand analysis based on activity models such as MATSim (Moeckel, 2018). For example, gravitation-based models date back to Carey in the 1850s, if not before, and are still widely used in transportation, land use, and migration analysis. This type of social physics was picked up by people like Stewart (1941) and found a clear expression in applications like the 1950s' Chicago Area Transportation Study (Voorhees, 1955) which marked the beginning of transportation-land use modeling. Tobler (1975) first developed cellular automata models in a spatial context, while spatial agent-based models were proposed in the 1950s by Hagerstrand (1968) and again in the 1960s by Chapin and Weiss (1968). Microsimulation dates from Orcutt's (1957) pioneering work in building economic models. Modern spatial statistics developed around questions of how spatial relationships impact the general linear model's assumptions particularly through spatial autocorrelation, which Moran (1950) first explored and whose specification in spatial statistics was formalized by Cliff and Ord (1973). A variety of space-time models emerged from this to form the basis for the new field of spatial econometrics (Anselin, 1988).

As urban analytics developed, the concept of the model—a formal representation of theory—has gained ground to the point where the term "theory" is now less-widely used than the term "model" (Batty, 2007). This is reflected in the dramatic shift toward induction and the prominence of data-driven machine learning methods. Some commentators, such as Anderson (2008), have controversially proclaimed "the end of theory," implying that the rise of big data and better models makes the traditional scientific method obsolete. Subsequent commentators however have critiqued the numerous epistemological problems with this stance. Atheoretical inductive searches for patterns in data have limited value because they rarely address the domain's big questions and open challenges, often misinterpret causal mechanisms and policy implications, and contribute little to science's fundamental theory-building endeavor. This is recognized in new models built on big data that utilize individual agent-based approaches, aiming for a new kind of social physics (Pentland, 2014).

Over time, the trajectory sketched here has seen models' elements—such as populations, economic activities, land uses, and geometric/topological features such as networks—disaggregated to introduce richer spatial and topical detail. Dynamic processes have been introduced to move beyond naïve equilibrium assumptions. Many of these developments require finer-scaled data and new kinds of data, much of which now come from real-time sensing and crowdsourcing. The models have grown richer but less parsimonious: new problems of fitting them properly to data have thus emerged.

In the following sections, we explore some of these new data sources and modeling methods, focusing on networks and big data. The earliest urban models established a concern for interactions and flows, though the networks that underpin such flows remained merely implicit, despite flows and networks being two sides of the same locational coin. One of the most exciting prospects for urban analytics is a stronger articulation of urban processes that take place on networks as flows, generating a new approach to locations as relations (Batty, 2013).

## Connections in Space

Urban analytics research and practice today enrich traditional spatial interaction theory with new methods from network science. Network science itself is introduced in detail in two other chapters here (Andris, this volume; Ye, this volume). This section places it in an urban context, then explores its current spatial formulation in urban analytics research and practice, focusing on networked urban infrastructure and transportation.

### Urban Spatial Networks

*Networks* are collections of entities connected to one another in some way. Network science builds on the mathematical machinery of graph theory to model network entities as *nodes* and their connections as *edges* in a mathematical representation called a *graph*. Directed graphs have edges that explicitly point one-way from an origin node to a destination node, while undirected graphs' edges point in both ways, bidirectionally. Common examples of real-world networks include the world wide web (a collection of web pages connected to one another by hyperlinks), social networks, and street networks. Real-world street networks are usually modeled as *primal graphs* with their intersections and dead-ends as nodes and their street segments as edges. Conversely, *space syntax* theory represents street networks as a *dual graph* where nodes represent named streets and edges represent their intersections (Jiang and Claramunt, 2002).

Spatial networks' nodes and/or edges are embedded in space. For example, a street network's intersections, dead-ends, and street segments exist in real geographical space and have corresponding geometric properties such as lengths, widths, areas, elevations, bearings, and angles. These explicitly spatial properties create unique opportunities—and constraints—for urban network analytics by intermingling spatial geometry with network topology. *Topology* refers to the configuration of the network's nodes and edges in relation to one another. For example, the distance between two nodes in a nonspatial graph is defined as the least number of edges that must be traversed between these two nodes, possibly weighted by each edge's strength of connection. This is a purely topological measure of distance. But a spatial network, such as a city's streets, merges geometry with topology. The distance between two points in the city is not merely topological: that is, it is not just the sum of edges traversed because these street segments have different impedances (e.g., length, traversal time, or elevation gain). Yet the distance is also not merely geometrical: human travel is constrained to transportation networks and is poorly estimated by straight-line distance. Instead, the distance between point A and point B is the impedance-weighted shortest path along the network edges.

Because urban infrastructure networks are spatially embedded, they are *approximately planar*, meaning that they can be generally well-modeled under the assumption that in two-dimensional space their edges intersect exclusively at nodes. Strict planarity breaks down in the presence of tunnels and bridges (Boeing, 2020b), but otherwise often approximately holds. Approximate planarity strongly constrains the density and connectivity of a spatial network. For instance, while a social network may include a node (i.e., person) with hundreds of incident edges (i.e., friendships), it is exceedingly rare for a street network to include a node (i.e., intersection) with more than 5 or 6 incident edges (i.e., streets) because of the approximate planarity constraint. Spatial embedding thus strongly influences street network topology and subsequent analytics. However, certain spatial networks, such as airports and airline routes, are not constrained by the planarity assumption because their edges are not embedded in two dimensions.

**Scientific Research on Urban Infrastructure Networks**

Spatial network science in urban analytics typically takes one of two general approaches depending on the research question: structural analysis or circulation analysis. Network *structural analysis* derives heavily from statistical physics and mathematics. It seeks to characterize the overall structure of the network, individual subcomponents or neighborhoods in the network, and the relative characteristics of an entity's position within the network. The past decade has seen rapid advances in our knowledge of urban infrastructure networks' clustering characteristics, distributions of centrality, and subcommunities and cliques. This research typically examines street networks as the archetype of networked urban infrastructure and emphasizes urban morphology—the study of the urban form's pattern and evolution. Such studies employ urban analytics inductively, examining data and attempting to make generalizable claims from their findings to advance urban science. This often results in abstract scientific knowledge rather than actionable insights and predictions for policymakers. However, recent advances in analyzing network resilience hold promise for better real-world urban planning.

The second key approach to spatial network science in urban analytics is *circulation analysis*, which has longstanding ties to planning practice. While structural analysis focuses on network patterns, configurations, and spatial signatures in the urban form, circulation analysis focuses on flows to analyze the movement of people, goods, capital, ideas, energy, water, and waste through a city's networked infrastructure. This can include the street networks previously discussed, but also airline networks, financial networks, telecommunication networks, and utility networks. This branch of analytics overlaps with qualitative approaches to urban studies, such as Castells' (2009) theory of global "spaces of flows"—which shape human society, culture, and economy independent of strong place ties—or Graham and Marvin's (2001) "splintering urbanism" theory, in which unequal provision of networked infrastructure shapes different subcommunities' interactions and experiences.

Circulation analysis in urban analytics relies on flow modeling. For example, the researcher constructs a model of the transportation network and then simulates trips along it based on origin/destination data, time of day, and mode of travel. These simulations use shortest-path algorithms with travel time impedances to solve routes and often integrate traffic assignment and queueing models. This can answer questions about daily travel patterns, infrastructure utilization, and urban growth. Accordingly, these network circulation analytics appear prominently in urban planning practice.

**Urban Network Analytics in Practice**

Large-scale urban simulation models have a mixed history of success and accuracy in technocratic central planning. For decades, metropolitan planning organizations have utilized integrated transportation-land use models to predict future urbanization, trip-taking patterns, and infrastructure needs (Waddell, 2011). They have also been used to simulate urban policy levers and estimate their effects. Traditionally, these analytics relied on spatial zones and Euclidean interactions but today planning practice increasingly operationalizes graph-based network models to analyze spatial connectivity.

Similar spatial network analytics in industry underpin tools like Google Maps and WalkScore, an app that quantifies pedestrian access to amenities in different neighborhoods. Spatial network models are also used for network-constrained spatial optimization problems (such as facility location) and GPS trajectory map-matching (such as snapping public transit vehicles to edges in a street graph, based on their device-reported locations). On one hand, these kinds of analytics help planners forecast urban development and travel, optimize infrastructure decisions under budgetary constraints, and lend an imprimatur of quantitative legitimacy to societal predictions. On the other hand, these predictions are only as good as their input data and model specifications, and thus sometimes lend unwarranted scientific credibility to what is nearly divination.

Spatial networks form an important component of urban analytics research and practice today. Compared to historical approaches to spatial interaction, recent advances in computational tractability allow urban analysts to better represent how urban systems are spatially configured and how people and resources actually move through space. In turn, structural analysis and circulation analysis of urban spatial networks have become key methods in the urban analytics toolkit. Such analytics today rely heavily on new sources of big data, which we turn to next.

# Spatiotemporal Big Data in Urban Analytics

**Definition, Characteristics, and Trends**

Since the earliest days of computing, computer scientists have wrestled with the concept of big data when dealing with data capture, storage, management, and processing (Kaisler, 2013). "Big data" refers to datasets too large to process with traditional databases, software, or hardware and collected from distributed sources or multiple disciplines. In 2001, Doug Laney coined the now-standard "3 Vs" of big data: volume (size or amount), velocity (speed of collection), and variety (source or format).

To address the challenges of managing and processing big data, researchers have extended this to include many more Vs, including veracity (truthfulness), value (usefulness), variability (temporal change), validity (accuracy), volatility (storage), vulnerability (security), visualization (presentation), etc.

Urban big data are big data that involve city activities and represent economic, social, cultural, and environmental aspects collected from sensors and devices embedded in cities. There is nothing inherently "urban" about big data itself, but in practice much big data is urban because of the enormous complexity of cities and increasingly ubiquitous urban sensing. The massive data exhaust of humans, natural and built environments, and their interactions in space and time are byproducts of the digital era. When combined with the right analytics and domain knowledge, they can help planners and policymakers derive new insights and solutions for major urban sustainability challenges in transportation, community development, design, housing, environmental justice, and public health, etc. A common refrain posits that data are the new oil. If so, then urban analytics is the engine producing data-driven policy and planning for the public realm. It applies suitable methods and tools to urban big data to derive novel urban insights and knowledge to enhance and support the policy- and decision-making for cities and communities to improve the quality of life for sustainable futures.

Applications of urban big data are distributed across a wide spectrum of data-as-a-service providers, aggregators, analytics consultants, and global technology giants. At one end of the spectrum, giants such as IBM, Cisco, Microsoft, Google, Amazon, and Alibaba have created sensors, networks, and platforms to collect, store, and manage urban big data and feed them into machine learning models for city government clients' smart cities initiatives. They focus on optimization and efficiency for city services, operations, and management—and are thus often critiqued by urban scholars and increasingly cities themselves for their narrow profit-driven interpretation of what residents want or need. At the other end, companies in specialized sectors such as mobility-as-a-service and other developers of smartphone applications have harnessed digital logs, trajectories, and transactions for millions of urban users. These user-generated troves of urban big data are usually sequestered in the hands of private companies, presenting barriers for both citizens and public servants to access and make full use of them for the public good.

Because of big data's complexity, researchers and scientists have developed a multitude of tools, models, and algorithms for data mining and knowledge discovery to extract insights. However, transforming urban big data into meaningful and actionable information for cities demands another level of effort. Urban big data are unique in their spatial, temporal, political, and ethical contexts and require both technical expertise and local domain knowledge. Here we briefly review several examples that address urban issues around the world at the vanguard of urban analytics. Due to urban analytics' interdisciplinary nature, many of these projects represent collaborations among planners, geographers, sociologists, economists, epidemiologists, computer scientists, engineers, physicists, and mathematicians.

**Urban Mobility and Its Impacts**

Mobility is perhaps the most widely studied example of urban big data due to ubiquitous availability and applicability. These analytics may or may not be constrained to spatial graphs, as discussed in the earlier section. Mobility data collection includes transactions and metadata from smart cards, smartphones, GPS devices, and intelligent transportation systems. These data all come with a timestamp and geolocation. The accuracy and resolution of such data vary depending on the sensing technology, hardware, and algorithms. For example, from cellular-based to triangulated call detail record data to GPS-enabled location-based smartphone app data, the spatial resolution may vary from a few kilometers in remote suburban areas to tens of meters in dense cities, and resolution will continue to improve along with technology.

Applying machine learning algorithms to individual mobility features, researchers have used metadata from millions of mobile devices to develop statistical models of individual activities at high spatiotemporal resolutions, yielding new travel demand estimates for metropolitan infrastructure planning (Jiang et al. 2016). Such models have been adapted to estimate electric vehicle charging demand and to measure building occupancy to estimate energy consumption. Combining mobile phone data with other data sources (such as air pollution, online housing transactions, and social media), researchers have studied the exposure of human beings to air pollution in the city, implications for environmental justice, and impacts on well-being and happiness (Nyhan et al., 2016; Xu et al., 2019; Zheng et al., 2019). Mobile phone data have been used by epidemiologists and urban scientists to study the effects of restricting human mobility to limit the spread of COVID-19 and to inform public health interventions in a number of countries (Bonaccorsi et al., 2020; Oliver et al., 2020; Zhou et al., 2020).

Using smart card transaction data, researchers can estimate travel demand and plan public transportation systems. With pick-up and drop-off data from transportation network companies, planners can optimize ridesharing networks and fleets (Santi et al., 2014). Emerging transportation modes such as bikesharing systems and electric scooters require further study to understand how active non-motorized transport systems serve residents and enhance sustainable urban mobility systems.

**Community Structure and Civic Services**

Social media data (such as check-ins, tweets, and crowdsourced reviews of businesses) are widely used to understand community structure, dynamics, and evolution. For example, the Livehoods project (Cranshaw et al., 2012) obtained 18 million location-based check-ins from Foursquare to study neighborhood boundaries by clustering urban inhabitants' activities. Using Twitter data, urban planners have applied natural language processing toolkits for sentiment analysis and other machine learning methods to measure perception, quality, and involvement of communities and neighborhoods over time (Schweitzer, 2014; Hollander et al., 2016; Williams, et al., 2019; Dong et al., 2019).

Other researchers have analyzed hundreds of millions of geocoded tweets and found persistent mobility isolation for economically disadvantaged residents beyond residential segregation (Wang et al., 2018). Offenhuber (2014) used 311 calls to understand the quality of urban services and legibility of local government, based on residents' complaints and feedback. These user-generated data allow us to examine human patterns and processes at unprecedented scale, but often exhibit biases by under-representing minority groups that do not use these platforms (Boeing, 2020a).

**Urban Design and Environment**

Google Street View images and photos voluntarily uploaded by individual users to social media and photo-sharing websites and have created new opportunities for planners and designers to evaluate the quality of urban space. Using photos posted on Flickr and Panoramio for hundreds of thousands of properties across the US, researchers found that the local volume of user-contributed, geotagged photos predicted ratings of the built environment's aesthetic quality (Saiz et al., 2018). Computer vision models and artificial intelligence techniques have been applied on images for various tasks related to urban design. Examples include predicting street safety scores to help neighborhoods improve urban physical environments (Naik et al., 2017), quantifying tree canopy coverage and relationships with public health (Li, et al. 2015), and deriving a vehicle profile census to explore voting patterns (Gebru et al., 2017).

Recent years have witnessed a rapid advance in spatiotemporal big data, sensor technology, and urban analytics applications. Yet this ubiquitous collection and analysis of human-centric data has grown increasingly contentious as pressing concerns about privacy, ethics, and power have emerged.

## The Dark Side of Urban Data

For decades, planners have used urban analytics to understand, monitor, predict, and control city spaces and the people who occupy them. These efforts have evolved rapidly in computational sophistication, spatially-explicit methodologies, and sources/volumes of data for model fitting. Yet as predictive power has improved, urgent new concerns have emerged around the privacy, ethical, and political implications of urban analytics and the often technocratic smart cities paradigm. Any accounting of the domain's city-making benefits must also grapple with its unprecedented, and often alarming, impacts on social life, particularly for vulnerable communities.

Privacy entails the ability of individuals to control information about their activities via space, time, and purpose. The Smart Cities technologies and data that underlie urban analytics extend surveillance, and thus portend a broadscale erosion of privacy and potential losses to due process. Such privacy losses, however, mean different things for different groups (Gilliom, 2001). Urban analytics alters virtually everything about government, corporate, and even family surveillance. These developments offer new opportunities to either extend or resist repression, depending on how legal and cultural contexts respond. The evidence so far suggests that treating new technologies and data as somehow socially "neutral" reinforces existing structures of domination, including racism and white supremacy (Noble, 2018).

A recent Google decision provides a ready example. The company received widespread criticism after refusing to remove a Saudi Arabian app called Absher from its online store. Absher is owned and operated by the kingdom's government and contains bureaucratic conveniences for complying with its male guardianship laws, including granting permission for a woman to take employment. Importantly, the app sends an alert to the owner whenever a woman they track uses a passport. It thereby potentially restricts travel and alerts violent domestic partners of a woman's attempt to flee. Domestic violence advocates objected that such an app violates women's rights to privacy, yet Google refused to remove the app, claiming it did not violate its terms of use. The terms of use did what the company needed: provide cover to avoid dealing with messy social issues around gender and justice. In so doing, the company reinforced a repressive status quo in what is ostensibly a simple governmental services app.

In turn, ethicists have failed to keep pace with the explosive growth of urban data and analytics. Standard theory traditionally held that individuals did not have privacy in public spaces like sidewalks or streets where they appear "in plain sight." In early discussions of technology and privacy, Nissenbaum (2004) argued instead that privacy ethics should center on "contextual integrity" that matches individuals' expectations of privacy with their context rather than strict rules about public and private settings. Today, Nissenbaum's premise erodes vis-à-vis Smart Cities and smartphone technologies of virtually all forms. Closed-circuit cameras have increased in resolution, ubiquity, and concealment. Image processing, storage, and analysis can be done with a fraction of the resources needed in prior decades. In addition to facial recognition, Smart Cities data can include information from many technologies like RFID transit passes, WiFi beacon sniffers, and automated license plate readers. Smartphones and any other wearable technology can track and send individual location data 24 hours a day, seven days a week. There is no context where individuals connected to technology are not generating data. In our world, humans are data-making machines, and many interests, both corporate and governmental, have a stake in these data.

Information asymmetries between data collectors, managers, and analysts ensure that individuals do not know what data they are generating, who has access to these data once they are collected, and how analytics related to these data structure the social, political, and economic choices that manifest. Consent to be surveilled becomes difficult to grant when so many avenues of surveillance exist and understanding them all requires extensive, specialized knowledge of both technology and corporate practices (Schweitzer and Afzalan, 2017).

Perhaps most importantly, passive sensors shift us away from legislated and (in the US) constitutional protections of due process toward new contexts where surveillance occurs by default. Wiretapping technology provides a useful comparison. To record conversations as evidence, due process required law enforcement to first obtain a warrant based on sufficient pre-existing evidence of wrong-doing. Certainly, law enforcement agencies have abused this process at times, but the default presumed that governments were not entitled to the information except when the specific situation warranted it. The burden of proof lay with the data collector before an individual, their activities, and their associates lost the presumptive right to privacy. But today, ubiquitous sensing places the burden on the individual to opt-out, when and if they have that option at all.

Cambridge Analytica's use of social media data in 2016 and prior elections to create effective, and readily manipulated, values profiles demonstrates how far analytics can go to alter the choices and opportunities individuals have, with potentially perverse results. Urban sensor data, especially combined with social media data, offer similar opportunities for manipulating individuals' choices and opportunities for association. While marketing and social groups have always influenced individuals' choices, corporate control over information of this type results from seemingly unrelated data collected then mass-aggregated for specific interests. Few avenues for accountability exist, and people do have things to fear from surveillance even if they are not doing anything wrong.

Rights to privacy and association, unfettered by structures of social control, are especially important to oppressed minorities. Limitless surveillance extends the possibilities for repression into every corner of the city where safe spaces once flourished. Urban policing technology and algorithms are foremost among the issues that could extend the reach and harm that local police already have on ethnic and racial minorities. Location data enhanced with sensor technology, let alone personal consumption data, can yield profiles that readily "out" those who engage in political protest or who gather in LGBTQIA venues to potential employers or police. Online doxxing, the release of an individuals' home or work address, takes on especially pernicious dimensions for people who live in cities surrounded by strangers who might attempt harm. This strategy, though ostensibly gender-neutral, often silences and punishes women who fail to cohere to the strictures of male domination.

Related to 24/7 surveillance problems, big data and social media can contribute to polluted or corrupted information environments in public policy and governance. Examples include "deep fake" videos of political opponents altered so they appear drunk or cognitively impaired. Certainly, crowdsourced videos have raised the potential to increase police accountability by challenging official accounts from multiple viewpoints. Nonetheless, in a corrupted information environment, evidence can become so eroded that deliberation—shared decision-making over a group of accepted facts—becomes much more time-consuming and challenging. It also raises a significant justice concern. Social, political, and economic power influence what people treat as "real" or "important" or "true" (Fricker, 2007). Rather than the information democracy for which technophiles yearn, information environments with infinite representations and known fakes potentially grant those with power even greater capacity to drown out or override contributions from those with smaller platforms and less institutional protection.

Despite the potential for repression, urban analytics and big data also have the potential to be powerful tools in promoting planning and public health. The same contact tracing that can "out" individuals and eliminates safe spaces was also one of the most useful tools that governments had in 2020 to counter the spread of COVID-19 (Kretzschmar et al., 2020). The same essential functions that apps like Absher use to track women can also help family members look after each other or enable women to find people nearby to help ally with them in unsafe situations. Urban analytics can either protect or imperil, subjugate or empower. The difference depends on social, political, and individual commitments to just practices as these tools become increasingly powerful.

## Towards a More Critical Urban Analytics

From its early roots in deductive science for urban prediction and planning, urban analytics has quickly grown to encompass emerging inductive methods from computer science and statistical physics. It enriches classical spatial interaction theory with new models and measures of urban connections and flows. It marshals big data to help planners understand urban mobility, community structure, and physical design. But do ubiquitous monitoring and modeling really help us solve the big, difficult problems dominating cities today?

How can urban analytics and big data help improve urban living and social justice, when they so often do not? What would a *critical* urban analytics look like, joining spatial data science with emancipatory and justice-seeking urban questions? Despite its promise of better city-making through better prediction, many applications of urban data and analytics entail alarming privacy and ethical consequences. But by learning from oppressed groups, a better critical approach to urban analytics may emerge.

For instance, members of vulnerable communities have leveraged urban analytics to resist domination—for example, by documenting and publishing information on supremacist extremists, locations of harassment and violence, and discriminatory employers and social organizations. As another example, compared to traditional customer service, automated systems improve certain urban service delivery. Young (2017) found that response times to service requests in low-income communities of color were faster under 311 app reporting and algorithmic assignment than those handled by public management staff. Brown (2019) found that app-based Uber and Lyft drivers racially discriminate against their passengers less than drivers for traditional taxi services do—though they do still discriminate. Sometimes more data and better algorithms can produce better outcomes for vulnerable communities.

An important and difficult challenge for urban analytics moving forward will be to harness and foreground these opportunities in investment and development efforts. Urban technologies and spatial analytics can empower repression or emancipation, largely determined by our governing practices. We must engage with the same critical choices as prior generations of analysis and information management. When information, whether created by astrology or sensor technologies, remains exclusively in elite hands, it produces power differences readily exploited for myriad ends, both good and ill. Urban analytics is no different.

A critical approach would intervene in knowledge creation systems to always question who benefits from new technologies and new data, and how those benefits might be extended or redirected equitably. This principle suggests straightforward methods for shifting the practice of urban analytics toward epistemic equality so that individuals know what data they generate, who uses it, and how they too might benefit from and control it. Such equality becomes possible only with a strong commitment to data literacy and open data systems. We see three critical fulcra: corporate governance, primary education, and professional education.

  First, companies should abide by stringent regulations and law on data protection and privacy (such as the EU General Data Protection Regulation). They should be required to report to individuals what data they have used and for what purposes, as well with whom they have shared or sold individuals' data. These reports will make the tradeoffs inherent in Smart Cities technology and ubiquitous sensing to empower urban analytics more legible to citizens. Public and private sectors should also work together to develop data-sharing guidelines to make anonymous big data available for urban analytics for the public good (Viggiano et al., 2020). Second, in the 21st century, it is no longer acceptable for policymakers or residents to be uninformed about data and technology. People should leave K-12 education with basic programming and data science skills along with a working understanding of how the Internet and sensors function to be informed voters, consumers, advocates, and citizen scientists. Third, urban planning, policy, and geography graduate education must prepare students to become leaders at the center of urban analytics and Smart Cities discussions. Too often, the social sciences cede this ground to physical scientists or engineers due to their emphasis on technical knowledge, but these discussions then lose critical perspectives and urban domain expertise as we have seen throughout the history of urban analytics.

  Sitting at the intersection of planning, spatial analysis, computer science, and statistics, the field of urban analytics provides a quantitative empirical foundation for researchers and practitioners seeking to understand, monitor, and improve cities. It has evolved in recent years from deductive, equilibrium-based models to inductive methodologies employing new sources of urban big data. But these data and attendant analytics entail serious consequences for representativeness, privacy, and equity. Today, an open challenge for urban analytics remains to move beyond the low-hanging fruit of analysis to contribute more critically to urban theory, policy-making, and the social equity and justice problems that afflict city planning. A crucial component of this is identifying the epistemological limits of urban analytics: what questions can it really answer, what problems can it really solve, and when should it defer to other forms of knowledge and ways of understanding the city?